\documentclass[aps,prl,twocolumn,superscriptaddress]{revtex4-1}
\usepackage{graphicx}

\begin{document}





\title{Ultrafast magnetic dynamics in insulating YBa${_2}$Cu${_3}$O$_{6.1}$ revealed by time resolved two-magnon Raman Scattering}


\author{Jhih-An Yang}
\author{Nicholas Pellatz}
\affiliation{Department of Physics, University of Colorado, Boulder, Colorado 80309, USA}

\author{Thomas Wolf}
\affiliation{Institut f{\"u}r Festk{\"o}rperphysik, Karlsruhe Institut f{\"u}r Technologie, P.O. Box 3640, D-76021 Karlsruhe, Germany}

\author{Rahul Nandkishore}
\affiliation{Department of Physics, University of Colorado, Boulder, Colorado 80309, USA}
\affiliation{Center for Theory of Quantum Matter, University of Colorado, Boulder, Colorado 80309, USA}

\author{Dmitry Reznik}
\email[Corresponding Author: ]{dmitry.reznik@colorado.edu}
\affiliation{Department of Physics, University of Colorado, Boulder, Colorado 80309, USA}
\affiliation{Center for Experiments on Quantum Materials, University of Colorado - Boulder, Boulder, Colorado, 80309, USA}



\begin{abstract}
Measurement and control of magnetic order and correlations in real time is a rapidly developing scientific area relevant for magnetic memory and spintronics. In these experiments an ultrashort laser pulse (pump) is first absorbed by excitations carrying electric dipole moment. These then give their energy to the magnetic subsystem monitored by a time-resolved probe. A lot of progress has been made in investigations of ferromagnets but antiferromagnets are more challenging. Here we introduce time-resolved two-magnon Raman scattering as a real time probe of magnetic correlations especially well-suited for antiferromagnets. Its application to the antiferromagnetic charge transfer insulator YBa$_2$Cu$_3$O$_{6.1}$ revealed rapid demagnetization within 90 fs of photoexcitation. The relaxation back to thermal equilibrium is characterized by much slower timescales. We interpret these results in terms of slow relaxation of the charge sector and rapid equilibration of the magnetic sector to a prethermal state characterized by parameters that change slowly as the charge sector relaxes. 

\end{abstract}



\maketitle


\section{Introduction}

Time-resolved studies of magnetism on ultrafast timescales span a wide range of materials from itinerant ferromagnets to insulating oxides \cite{RevModPhys.82.2731}. However, they have many limitations and there is a great need to develop new techniques especially for antiferromagnets (AFs). Time-resolved magneto-optical Kerr Effect (MOKE) \cite{PhysRevX.9.021020}, circular dichroism \cite{boeglin2010distinguishing}, or time-resolved angle-resolved photoemission (trARPES) \cite{tengdin2018critical} see ferromagnetic (FM) ordered moment and can only be applied to ferromagnets and canted antiferromagnets. Optical measurements have been used extensively to investigate AFs, but they provide only indirect information about the magnetic channel \cite{PhysRevB.82.060513,PhysRevB.83.125102,ncomms6112,Miyamoto2018}. Stimulated Raman scattering works only in special cases and is also indirect \cite{batignani2015probing}. Resonant x-ray diffraction is a powerful probe of magnetic order \cite{ehrke2011photoinduced,zhou2014glass,chuang2013real,caviglia2013photoinduced,lee2012phase} but in many materials (such as cuprates), kinematics do not allow access to AF zone boundary ordering wavevectors. Resonant inelastic x-ray scattering (RIXS) has been used once in the time-resolved mode, but neither time nor energy resolution were sufficient to see strong effects in the magnon spectrum \cite{nmat4641}.

\begin{figure*}
\includegraphics[trim={0cm 0cm 0 2cm},width=\textwidth]{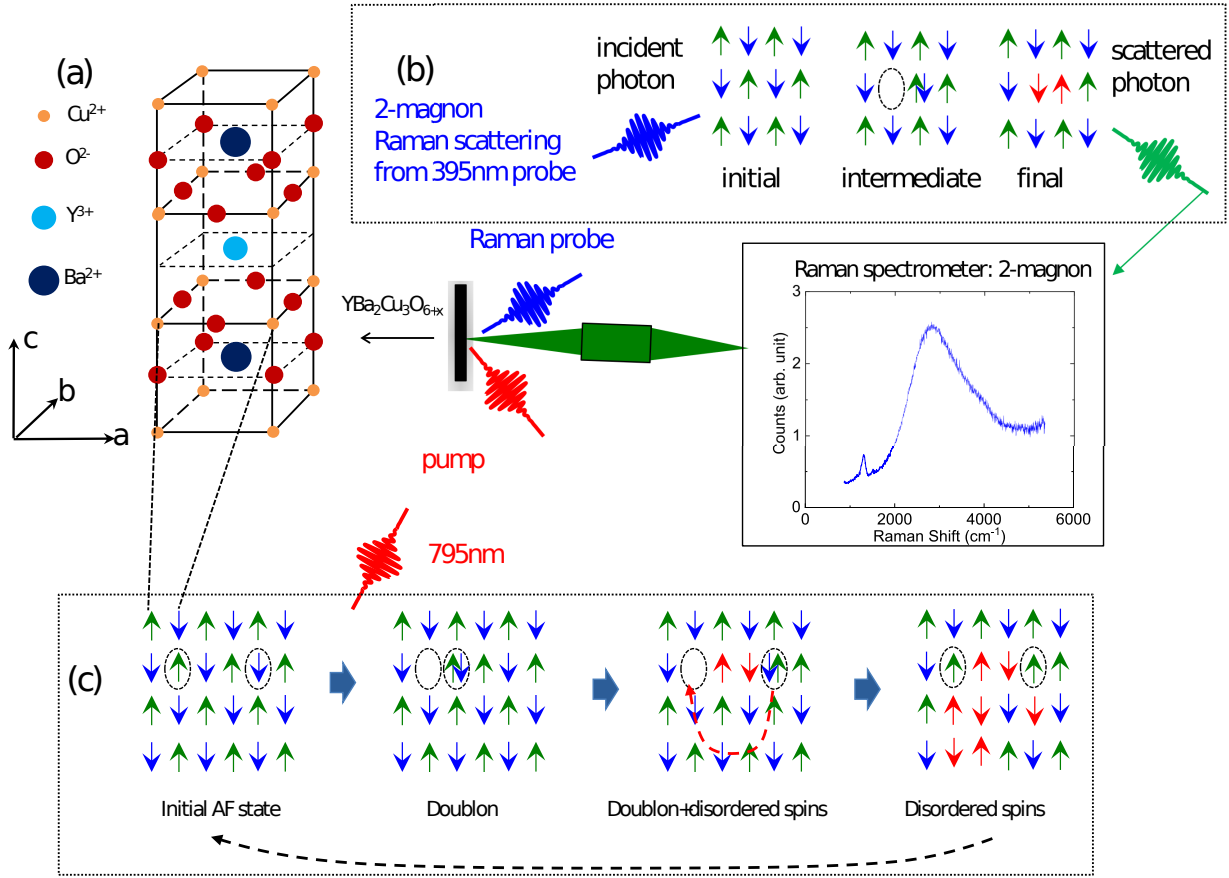} 
\caption{Schematic of the pump-probe two-magnon Raman scattering experiment on YBCO. (a) Structure of YBCO. (b) A cartoon of the 2M Raman process in an antiferromagnet used as a probe. The sample is hit with the (blue) probe pulse. Raman-scattered light (green) is collected into the spectrometer. The broad peak with a maximum at 2700 $\textrm{cm}^{-1}$ shown in the window is the 2M scattering signal obtained with the 532 nm laser. The mechanism of the 2M scattering process is illustrated in the schematic above the data window: An electron is hopping to its nearest neighbor by absorbing an incoming photon, creating an unoccupied site and a doubly occupied site. The other electron in the doubly occupied site with opposite spin hops back to the hole site with the emission of an outgoing photon. The final state has six exchange bonds that are energetically unfavorable. (c) Schematic of photoexcitation followed by relaxation as discussed in the text: Creation of holons/doublons followed their diffusion and recombination accompanied by spin flips, and finally, relaxation of magnetic excited state to thermal equilibrium.
\label{Schematic}}
\end{figure*} 

Two-magnon (2M) Raman scattering from ultrafast laser pulses used as a probe fills this gap. We developed it while investigating YBa$_2$Cu$_3$O$_{6.1}$ (YBCO), which is of broad interest as an AF charge transfer (CT) insulator whose hole doping leads to high temperature superconductivity (Fig. \ref{Schematic}a). CT and similar Mott/Hubbard insulators are characterized by a large Coulomb energy cost for an electron to hop onto a site that is already occupied by another electron. This interaction splits the conduction band into the lower and upper Hubbard bands, which correspond to singly and doubly occupied sites. Typically these bands are separated by a gap, $U$, of the order of 2 eV. In insulating phases the conduction band is half-filled (one electron per site), electrons are localized, and their magnetic moments order antiferromagnetically. 

It is well-known from literature that the Raman spectrum of insulating oxide perovskites is dominated by a broad intense two-magnon (2M) Raman peak \cite{PhysRevB.3.1709,PhysRevB.39.9693,PhysRevB.38.6436,RevModPhys.79.175}. In YBCO its maximum is at 2720 $\textrm{cm}^{-1}$ (Fig. \ref{Schematic}b) \cite{Ponosov1990551,PhysRevB.48.7624,PhysRevB.97.024407}. Fig. \ref{Schematic}b illustrates the schematic of the 2M scattering process that is responsible for the excitations giving rise to the 2M peak maximum: the incident laser photon is absorbed by a virtual creation of a doubly-occupied site and an empty site. Then the other electron jumps to fill the hole ending up with two near neighbor moments flipped. The energy cost due to the breaking of six exchange bonds is about 3$J$. In other words, the 2M peak mainly originates from slowly dispersing magnons near the zone boundary, which corresponds to short-wavelength magnons. As a result, 2M Raman scattering is sensitive to short-range AF order and is a powerful tool for the study of magnetic excitations and correlations. It has been reported in many Mott and charge transfer (CT) insulators such as YBCO. 

Energy injected by pump laser pulses can be efficiently absorbed by electrons hopping into doubly-occupied neighboring sites when the pump photon energy is equal to or greater than $U$ (left two diagrams in Fig. \ref{Schematic}c). The resulting doubly-occupied sites (doublons) and unoccupied sites (holons) both have zero magnetic moment. They disorder remaining magnetic moments as they diffuse through the atomic lattice (diagrams on the right in Fig. \ref{Schematic}c). The resulting demagnetization should impact the magnon spectra and, as a consequence, two-magnon Raman scattering. In particular, scattering intensity near the peak originates from antiferromagnetically ordered regions without holons or doublons, so it is a good measure of the underlying short range magnetic order.  Here, we present direct evidence that optical pumping does in fact lead to a rapid suppression of the 2M Raman scattering, indicating demagnetization.  We show how the timescales on which the demagnetization and recovery occur as well as a phenomenological analysis of the lineshape of the 2M peak can help illuminate the interaction of the electronic and magnetic sectors.

\section{Results}
\subsection{Experimental details}
The 2M peak appears in the $B_{\textrm{1g}}$ symmetry, which can be isolated by polarization analysis of incident and scattered probe pulses. The configuration $xx$/$xy$ indicates that the incident laser polarization is parallel to the primitive cell in-plane crystal axes ($a$ and $b$), and the scattered light polarization is parallel/perpendicular to the incident laser polarization respectively. The $x'$ and $y'$ directions are rotated 45$^{\circ}$ in the $ab$-plane with respect to $a$ and $b$. $xx$, $xy$, $x'x'$, and $x'y'$ polarization geometries measure $A_{\textrm{1g}}$ + $B_{\textrm{1g}}$, $A_{\textrm{2g}}$ + $B_{\textrm{2g}}$, $A_{\textrm{1g}}$ + $B_{\textrm{2g}}$, and $A_{\textrm{2g}}$ + $B_{\textrm{1g}}$ symmetry components, respectively, in the $D_{\textrm{4h}}$ point group.

Traditionally the 2M Raman spectra are obtained with CW lasers. We added time domain to these measurements by using a system based on an amplified mode-locked 20 kHz Ti:sapphire laser \cite{Yang2017graphite}. Its fundamental 790 nm ($\approx$1.5 eV) laser pulses, whose photon energy happens to be very close to the YBCO CT gap, were used as the pump. The pulse at 395 nm from second harmonic generation (SHG) was used as the probe. Its photon energy resonates with the virtual absorption (transition from the initial state to intermediate state in Fig. \ref{Schematic}b) in the 2M scattering process around 3 eV \cite{PhysRevB.53.R11930}, which enhances the signal. All optics after SHG including focusing of the lasers onto the sample were mirror-based to ensure high time resolution. Measurements were performed at room temperature on a McPherson triple Raman spectrometer equipped with the liquid nitrogen-cooled CCD detector. It was calibrated with a standard McPherson Model 621 light source supplied with CSTM-CAL-LX-IRR calibration certificate.  Our experiments did not probe phonons as the large width of the laser line in energy necessary to obtain good time resolution precludes measurements below 800 cm$^{-1}$, which is above the phonon energies. Phonon measurements requiring significant modifications of the experimental apparatus are planned for the future. 

Since we are interested only in the Raman scattering from the probe laser pulse, photon counts obtained with the pump alone were considered as background and subtracted from raw data. This background increases with increasing pump power, with statistical uncertainly also increasing as a result.

Single crystals of YBa$_2$Cu$_3$O$_{6+x}$ were grown as described in \cite{wolf89} except that zirconia crucibles were used. To characterize the sample we measured Raman scattering from phonons with high resolution using the CW 532 nm diode laser. The phonon spectrum, which is very sensitive to doping, was consistent with the oxygen concentration well below the metal-insulator transition (close to $x = 0.1$). Measurements were performed at room temperature with the sample in air. Time-average laser powers used were around 3 mW for the probe and 10-30 mW for the pump. Based on these values and on past experience, we estimate no more that 10 K of average heating by the lasers. Self-pumping from the probe laser alone was much smaller than the effect of the pump (see Supplementary Note 1).

\begin{figure}
\centering
\includegraphics[trim={0cm 0cm 0 0cm},clip,width=0.5\textwidth]{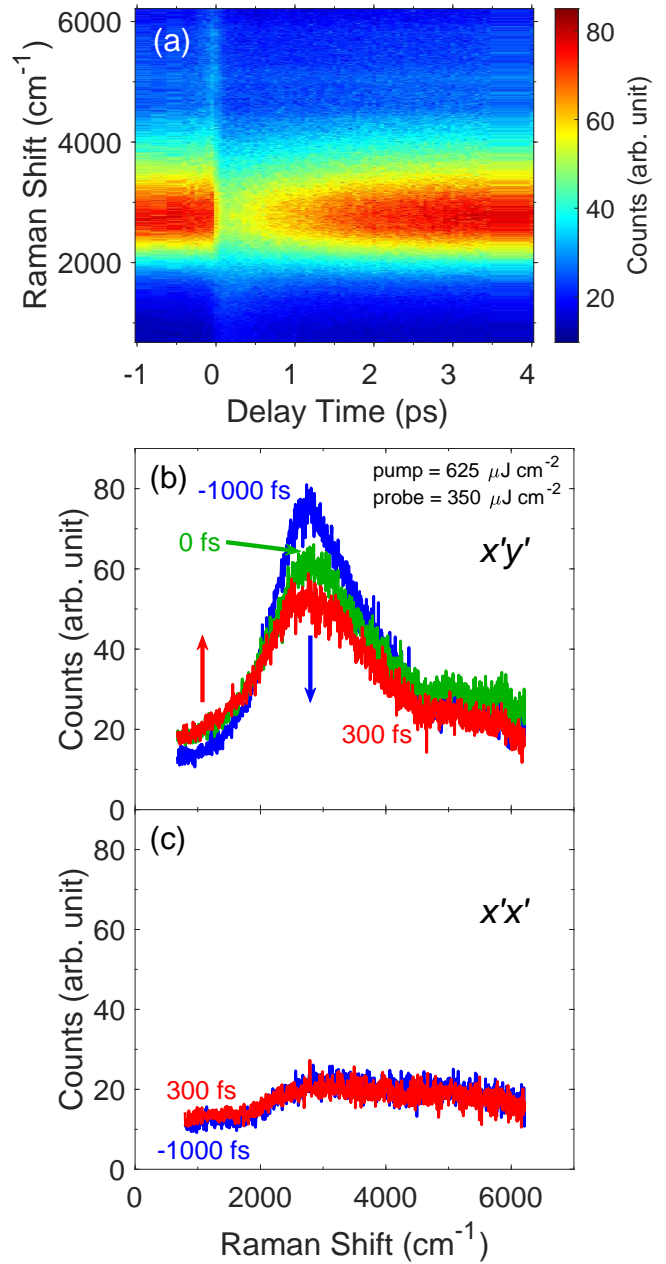} 
\caption{Time-resolved two-magnon Raman scattering in YBCO. (a) Colormap representing two-magnon peak as a function of time after photoexcitation. Negative times correspond to time before photoexcitation. (b,c) Raw data at several time delays in the $x'y'$ (b) and $x'x'$ (c) geometries.
\label{Raw_Data}}
\end{figure}

\subsection{Time-dependence of two-magnon scattering}

\begin{figure}
\centering
\includegraphics[trim={0cm 0cm 0cm 0cm},clip,width=0.45\textwidth]{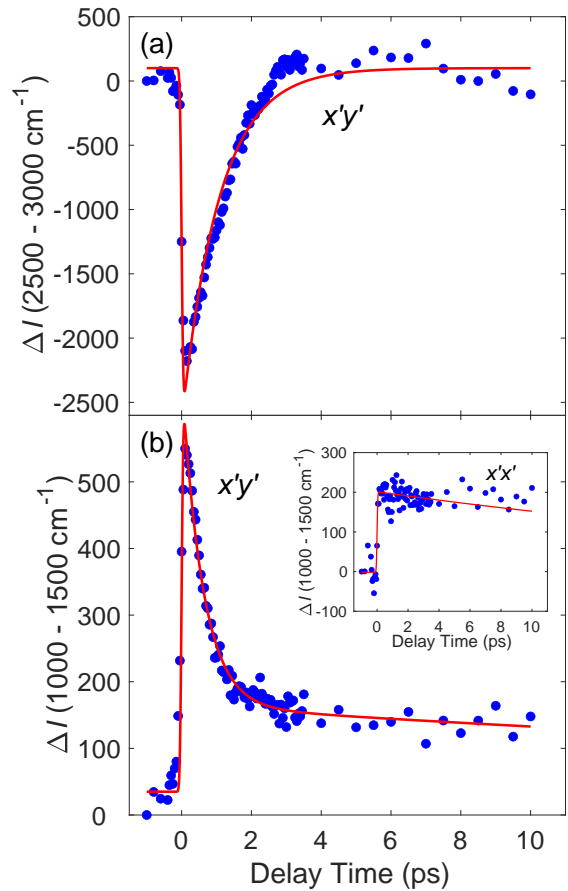} 
\caption{Integrated intensity in Fig. \ref{Raw_Data} from 2500 $\textrm{cm}^{-1}$ to 3000 $\textrm{cm}^{-1}$ covering the main $x'y'$ two-magnon peak (a) and from 1000 $\textrm{cm}^{-1}$ to 1500 $\textrm{cm}^{-1}$ (b) at various delay times, relative to the intensity at $t = -1000$ fs (labelled $\Delta I$). Inset shows integrated intensity in the $x'x'$ geometry from 1000 $\textrm{cm}^{-1}$ to 1500 $\textrm{cm}^{-1}$. The solid lines are fits to a biexponential function. The time constant in (a) is 1040 fs. In (b) the fast decay time constant is 675 fs and the slow decay constant is over 50 ps for the inset. The noise is mainly due to fluctuations in beam pointing.
\label{2M_time}}
\end{figure}

Figure \ref{Raw_Data} shows that the 2M peak is suppressed after an ultrafast optical excitation (indicated by a blue arrow) with some spectral weight moving below 2000 $\textrm{cm}^{-1}$ (indicated by a red arrow). The effect is present only in the $x'y'$ geometry and not in the $x'x'$ geometry, consistent with the symmetry analysis. Time-dependence of the scattering intensity integrated from 2500 $\textrm{cm}^{-1}$ to 3000 $\textrm{cm}^{-1}$ and from 1000 $\textrm{cm}^{-1}$ to 1500 $\textrm{cm}^{-1}$ (Fig. \ref{2M_time}) gives a quantitative picture. The initial jump/drop of the intensity occurs within our time resolution of 90 fs deduced from a cross-correlation measurement. We never saw any measurable signal on the anti-Stokes side of the spectrum.

\begin{figure}
\centering
\includegraphics[trim={0cm 1cm 0cm 0cm},clip,width=0.5\textwidth]{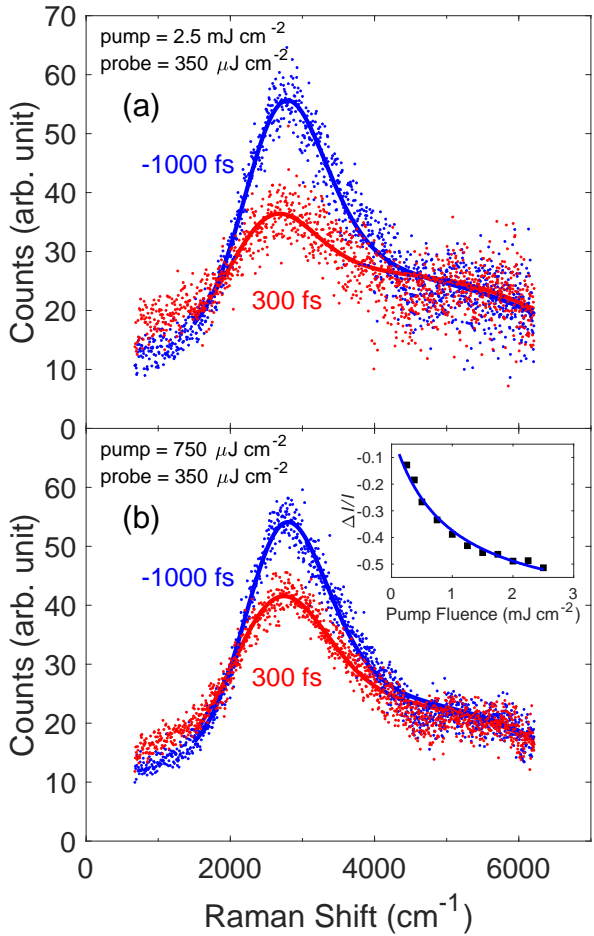} 
\caption[Fluence dependence of magnetic melting of YBCO]{Fluence dependence of magnetic melting of YBCO. (a,b) Two-magnon Raman spectra at $t = -1000$ fs and $t = 300$ fs with pump fluence of 2.5 mJ cm$^{-2}$ (top) and 0.75 mJ cm$^{-2}$ (bottom) in the $x'y'$ configuration. Luminescence from the pump only was subtracted. Inset: The pump fluence dependence of integrated intensity change from $t = -1000$ fs to $t = 300$ fs. $\Delta I/I$ is the integrated intensity from 2200 $\textrm{cm}^{-1}$ to 3400 $\textrm{cm}^{-1}$  normalized to the integrated intensity at $t = -1000$ fs. The background that does not originate from two-magnon scattering was subtracted to determine the extent of magnetic melting (this background was measured in the $x'x'$ configuration).  We observe a trend towards saturation at high pump fluence. The solid curve is a fit to a saturation model $F/(1+F/F_0)$ with $F_0=0.9$ mJ cm$^{-2}$. 
\label{2M_power}}
\end{figure}

Figure \ref{2M_power}a,b shows that the reduction of the peak intensity of the $x'y'$ spectra with pump fluence of 2.5 mJ cm$^{-2}$ is a factor of 2 greater than for a much smaller pump fluence of 0.75 mJ cm$^{-2}$.  The time scale of the 2M recovery ($\sim$1 ps) is similar at different pump powers, but the magnitude of melting follows a nonlinear power dependence (Figure \ref{2M_power} inset). We fit this fluence dependence with a saturation model $\sim F/(1+F/F_0)$ with the fitting parameter $F_0=0.9$ mJ cm$^{-2}$. 

After reaching a minimum around $t = 100$ fs the integrated intensity from 2500 $\textrm{cm}^{-1}$ to 3000 $\textrm{cm}^{-1}$ recovers following an exponential function with a time constant of 1040 fs (Fig. \ref{2M_time}a). Small systematic deviations from the fitted line are consistent with beam pointing fluctuations. 

Figure \ref{2M_time}b shows the dynamics of $x'y'$ low energy excitations (integrated from 1000 $\textrm{cm}^{-1}$ to 1500 $\textrm{cm}^{-1}$) fit with a biexponential function. We use biexponential function because the scattering intensity shows a fast decay at small times (time constant of 675 fs) followed by very slow decay. The latter does not come back to the intensity before photoexcitation (at negative time) even at 10 ps. We place its lower bound around 50 fs based on fits to the available data. The same energy interval in the $x'x'$ configuration exhibits the slow decay only (see Fig. \ref{2M_time}b inset). 

\begin{figure}
\centering
\includegraphics[trim={0cm 0cm 0cm 0.6cm},clip,width=0.45\textwidth]{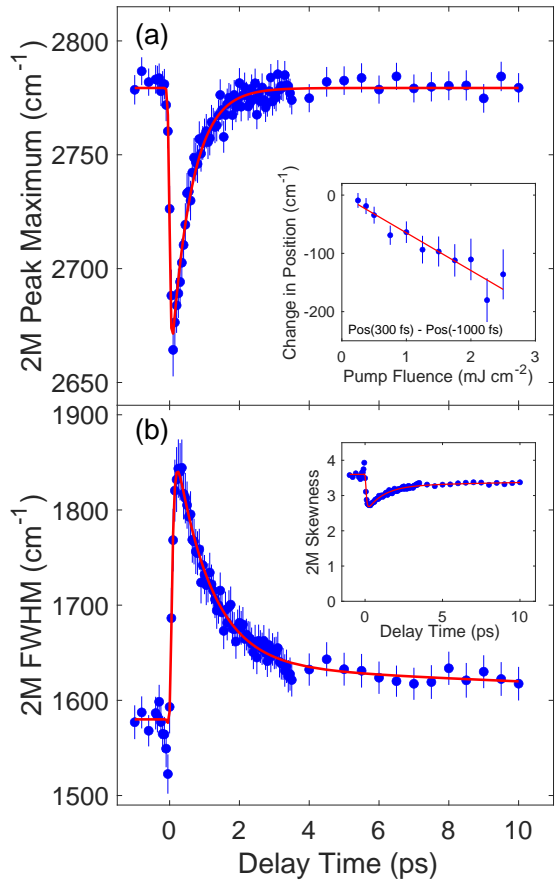} 
\caption{2M peak position and linewidth as a function of time for the fluence of 625 mJ cm$^{-2}$. (a) 2M peak position; The solid line is a fit to a single exponential function with the decay constant of 589$\pm 44$ fs. Inset shows peak position shift from $t = -1000$ fs to $t = 300$ fs as a function of fluence; (b) 2M peak width. The fit is to a biexponential function with a fast decay of 1003$\pm 134$ fs and slow decay of about 30 ps. The skewness shown in the inset (see Supplementary Note 2 for definition) exhibits similar time dependence to the peak width.  Error bars for all quantities are 95\% confidence intervals calculated from the fitting procedure described in the Supplementary Note 2.
\label{2M_peak}}
\end{figure}

The lineshape of the 2M peak is determined by many factors in addition to the magnon joint density of states including magnon-magnon interactions, electron-photon matrix elements, and spin-phonon coupling. It is particularly important to emphasize here that 2M Raman scattering is a strongly resonant process, so the exact lineshape depends not only on the magnon spectrum, but also on interactions with high energy electronic excitations on the order of the probe laser energy. Relevant theory has not been developed for the nonequilibrium regime, thus extracting more detailed information from lineshape analysis is beyond the scope of this work. On a purely qualitative level the peak is asymmetric with more spectral weight on the high-energy side. This asymmetry was explained as originating from coupling to an apical oxygen phonon \cite{farina2018electron}. We fit the 2M peak using a phenomenological lineshape that is essentially a Gaussian peak, modified to be asymmetric, on top of a background as explained in detail in Supplementary Note 2 (Fig. \ref{2M_power}). The peak maximum softens by a maximum of 4\% when the pump fluence is 625 mJ cm$^{-2}$ as a consequence of photoexcitation and comes back within 600 fs (Fig. \ref{2M_peak}a). Increase in the pump fluence further softens the 2M peak as shown in the inset to Fig. \ref{2M_peak}a. Meanwhile the linewidth increases at short times and appears to be peaked around 300 fs (Fig. \ref{2M_peak}b). The asymmetry of the peak drops initially and then recovers similarly to the linewidth (see Fig.  \ref{2M_peak}b, inset). 

\section{Discussion}

In this section we discuss the interpretation of the results. We argue that all the results can be explained within a simple picture whereby the charge sector relaxes slowly (with a timescale of $\sim$1 ps, which can be understood in terms of the standard theory of prethermalization), whereas the magnetic sector relaxes rapidly (i.e. with a timescale shorter than experimental resolution), and can be viewed as being in instantaneous equilibrium at a given configuration of the charge sector. Theoretical considerations suggest that the dominant relaxation channel for the charge sector involves multimagnon emission, and this scenario is consistent with the data, however the interpretation is insensitive to the precise nature of the dominant relaxation channel, and clarifying the role of phonons (if any) remains an interesting topic for future work. 

We begin by discussing the nature of the absorption (Fig. 1) and the suppression of the 2M peak (Fig. 2). The pump photon energy matches the onset of absorption in optical spectra, i.e. it corresponds to holon-doublon pair creation energy. Therefore it is likely that holon-doublon pair creation dominates absorption of pump photons \cite{ncomms6112}. However, the timescale for pair creation is evidently less than our $90$ fs time resolution. Since holon-doublon pairs are optically-active, it is instructive to compare relaxation of the two-magon peak with the results of transient optical conductivity \cite{PhysRevB.82.060513,PhysRevB.83.125102, ncomms6112}. These show that pump pulse first generates a metallic Drude-like spectrum, which indicates that photoinduced doublons and holons (Fig. \ref{Schematic}c) are mobile right after photoexcitation. As a doublon or holon hops through the atomic lattice, it leaves behind a trail of flipped spins (Fig. \ref{Schematic}c). Energy release from holon-doublon pair recombination further disrupts the AF order and, as a consequence, suppresses the 2M peak (Fig. \ref{Raw_Data}). 

We now discuss relaxation of the charge (holon-doublon) sector through recombination. Recombination releases energy, which must be transferred to other excitations. The obvious candidates are either phonons or magnons. However, the energy release through recombination is of order $U$, which far exceeds the local bandwidth in both the magnon and phonon sectors. Recombination must thus involve the emission of multiple phonons or magnons, and may be described by the well developed theory of prethermalization \cite{RoschPrethermal, MoriPrethermal, AbaninPrethermal}. We present here a simple version of the argument, following \cite{RoschPrethermal}. For decay to a sector with maximum local bandwidth $\Lambda$, the recombination must involve emission of at least $U/\Lambda$ excitations, and thus can only occur at order $U/\Lambda$ in perturbation theory. The matrix element is thus of order $U (\Lambda/U)^{U/\Lambda}$, which is exponentially small in $U/\Lambda$. This calculation can be made rigorous \cite{MoriPrethermal, AbaninPrethermal} yielding a relaxation rate $\Lambda \exp(- \alpha U/\Lambda)$ and hence a relaxation time $\Lambda^{-1} \exp( \alpha U/\Lambda)$, where $\alpha$ is an $O(1)$ constant. 

The local bandwidth for one-phonon excitations is of order 80 meV \cite{Pintschovius2004}, whereas the local bandwidth for one magnon excitations is of order 250 meV \cite{PhysRevB.54.R6905}. Since the relaxation rate is {\it exponentially} sensitive to the local bandwidth, it is clear that the magnon sector, having larger local bandwidth, should provide the dominant channel for relaxation. We therefore ignore phonons henceforth, and assume that recombination proceeds through multimagnon emission. We will show that this assumption provides a consistent interpretation of the data. However, it is not a crucial assumption, since analogous arguments could be made if multiphonon emission was dominant instead, although in that case we would need to make one additional assumption regarding rapidity of equilibration timescales within the magnetic sector. Nonetheless, clarifying the role of phonons (if any) remains an important topic for future work. 

In terms of experimental evidence for phonons vs magnons, we note, that the dominance of multimagnon emission is consistent with the observation that carrier relaxation time scale in semiconductors (see e.g. \cite{PhysRevB.88.165313emicon}), in which there are no magnons, is about 2-3 orders of magnitude longer than that observed in itinerant (metallic) ferromagnets (with magnons) where the demagnetization time scale is around 100-200 fs followed by $\sim$1 ps recovery \cite{PhysRevLett.76.4250,tengdin2018critical}. However, there are experiments (on different cuprates) that see relaxation to phonons \cite{DalConte1600}, and an unambiguous determination of whether phonons or magnons are the dominant channel would be an interesting problem for future work. For the present, we assume multimagnon emission is dominant, and this allows us to cleanly explain our results.  


We now discuss timescales for relaxation through multimagnon emission.  It was estimated in \cite{PhysRevLett.111.016401} that $U/J \ge 5$, and hence  recombination must be accompanied by the emission of at least 5 magnons. The calculation by Lenarcic et al. \cite{PhysRevLett.111.016401} further obtained a numerical estimate for $\alpha$, yielding a relaxation time of 90 fs for Nd$_2$CuO$_4$. However, it is important to note that this numerical estimate has an unknown uncertainty associated with it, especially since the calculation was done for a related but substantially different material, and since $\alpha$ sits in the exponential, this uncertainty in $\alpha$ translates into a much larger uncertainty in the relaxation timescale. 
We conjecture that the $\sim$1 ps fast decay timescale observed in our experiments corresponds to this prethermal timescale for holon-doublon recombination. On timescales short compared to this timescale, the holon-doublon concentration may be viewed as an effectively conserved quantity. 

We now discuss equilibration within the magnon sector. Within the magnon sector there is no relaxation bottleneck analogous to the recombination energy, and indeed one magnon hopping (matrix element $\sim J$) and two magnon interactions (matrix element $\sim J^2/U$) should be sufficient to produce thermalization. Since multimagnon processes are not necessary for equilibration within the magnetic sector, the magnetic sector should equilibrate much faster than the charge sector. If the charge sector equilibration time is $\sim$1 ps, and the magnetic sector equilibrates much faster, then the thermalization time for the magnetic sector is likely less than our temporal resolution of 90 fs. Accordingly, we propose that we should interpret our results in terms of a magnetic sector that is in equilibrium at an effectively fixed density of holons and doublons. 

This provides a simple interpretation of our results. The fast relaxation timescale of $\sim$1 ps (Fig. \ref{2M_time}) is the relaxation timescale for the holon-doublon sector. On timescales short compared to this, the holon-doublon concentration is approximately conserved, however the magnon sector is in effective equilibrium at a doping level determined by the doublon-holon concentration. It is well known that the effective $J$ is renormalized down (i.e. reduced) by doping \cite{mentink2017manipulating} and this provides a natural explanation for the observed mode softening (Fig. \ref{2M_peak}). The pump effectively photo-dopes the sample, which renormalizes $J$ down, leading to mode softening. A larger pump fluence produces more holon-doublon pairs, and hence a greater effective doping at short times, and thus produces a larger renormalization of $J$ (i.e. a more pronounced mode softening). The photo-doping is spatially non-uniform, and so there is some spatial heterogeneity to the renormalization of $J$, which broadens the peak. On timescales long compared to the holon-doublon relaxation time, the doping disappears, and so does the mode softening and the peak broadening.


There remains to explain only the slow decay with a time constant of around 50 ps in Fig.\ref{2M_peak} (with the caveat that these large relaxation times are obtained from a fit over a 10 ps time window). Such slow decays are also observed in optical measurements \cite{PhysRevB.50.4097,PhysRevB.82.060513,PhysRevB.83.125102}. Its appearance in both $x'y'$ and $x'x'$ configurations points at electronic Raman scattering from trapped charge carriers. It is known that after most holon-doublon pairs recombine, the remaining carriers form localized mid-gap states observed by optical probes \cite{ncomms6112,PhysRevB.82.060513,PhysRevB.83.125102,PhysRevB.89.165118}. These form and decay much more slowly. It was proposed that these states form polarons with either spin or phonon bosonic field component. Ultrafast suppression of the 2M peak intensity in our experiments confirms that mid-gap states perturb the AF order, and we conjecture that these slowly decaying mid gap states are responsible for the slow decay that we observe.

Since 2M Raman scattering only probes short-range spin correlations in the 2D copper oxide planes, it would be desirable to figure out what happens to the 3D magnetic order after an intense optical excitation. Due to a much smaller inter-layer exchange constant, we expect the recovery of 3D order to be slower than of the 2D order. Although a similar scenario has been observed in Sr$_2$IrO$_4$ \cite{nmat4641}, more experiments and calculations are needed.

Our experiments demonstrated that time-resolved 2M Raman scattering is a powerful probe of ultrafast demagnetization in antiferromagnetic Mott/charge transfer insulators. It can be applied to a variety of materials where two-magnon Raman scattering has been observed \cite{RevModPhys.79.175}. Electron-spin coupling plays an important role in photo-carrier relaxation as evidenced by a radical disturbance of the magnon spectrum. In YBCO the maximum effect occurred within our experimental resolution of 90 fs of photoexcitation, which is much faster than demagnetization timescales in itinerant ferromagnets \cite{tengdin2018critical}. Slower timescales characterizing subsequent relaxation to thermal equilibrium are of the same order of magnitude or greater than phonon-driven relaxation in conventional materials (e.g. graphite) \cite{PhysRevB.80.121403,Yang2017graphite}. We have proposed a simple explanation of these results in terms of slow relaxation of the charge sector and fast relaxation of the magnetic sector. Our results demonstrate strong coupling between charge and spin degrees of freedom, which potentially accounts for high-$T_\textrm{c}$ superconductivity. The coupling between spin and phonon degrees of freedom remains to be understood. In the future tuning pump laser energy to resonate with particular dipole-active phonons may allow using these measurements to provide new insights.

\section{Data Availability}
The datasets generated during and/or analyzed during the current study are available from the corresponding author on reasonable request.

\section{References}


\section{Acknowledgements}
Work at the University of Colorado was supported by the NSF under Grant No. DMR-1709946 (J.-A.Y., N.P., D.R.) and by DARPA through the DRINQS program (R.N.) We thank Justin Griffiths and Kyle Gordon for help with the experiments. 

\section{Author Contributions}
J.-A.Y. performed measurements and analyzed data, N.P. analyzed data, and worked on the manuscript, Th. Wolf grew the sample, R.N. provided theory support and edited the manuscript, D.R. conceived and oversaw the project, wrote and edited the manuscript.

\section{Competing Interests}
The authors declare no competing interests.

\end{document}